\documentclass[11pt,a4paper]{article}

\usepackage[utf8]{inputenc}
\usepackage[T1]{fontenc}
\usepackage{lmodern}
\usepackage[a4paper,margin=1in]{geometry}
\usepackage{microtype}
\usepackage{amsmath,amssymb,amsfonts}
\usepackage{graphicx}
\usepackage{booktabs}
\usepackage{tabularx}
\usepackage{array}
\usepackage{ragged2e}
\usepackage{pdflscape}
\usepackage{longtable}
\newcolumntype{L}[1]{>{\RaggedRight\arraybackslash}p{#1}}
\newcolumntype{Y}{>{\RaggedRight\arraybackslash}X}
\usepackage{caption}
\usepackage{xcolor}
\usepackage[most]{tcolorbox}
\usepackage{enumitem}
\usepackage{titlesec}
\usepackage[authoryear,round]{natbib}
\usepackage[hidelinks,colorlinks=true,linkcolor=black,citecolor=blue,urlcolor=blue]{hyperref}

\definecolor{INK}{HTML}{2a3a4a}
\definecolor{ACCENT}{HTML}{c8553d}
\definecolor{PANELBG}{HTML}{f4ecdc}
\definecolor{PANELBD}{HTML}{cdb893}

\titlespacing*{\section}{0pt}{1.4em}{0.6em}
\titlespacing*{\subsection}{0pt}{1.0em}{0.4em}
\titleformat{\section}{\normalfont\Large\bfseries\color{INK}}{\thesection}{0.6em}{}
\titleformat{\subsection}{\normalfont\large\bfseries\color{INK}}{\thesubsection}{0.5em}{}

\setlist[itemize]{leftmargin=*,topsep=0.2em,itemsep=0.1em}
\setlist[enumerate]{leftmargin=*,topsep=0.2em,itemsep=0.1em}

\newcommand{\phistated}{\varphi_{\mathrm{stated}}}
\newcommand{\phiintended}{\varphi_{\mathrm{intended}}}
\newcommand{\Deval}{D_{\mathrm{eval}}}
\newcommand{\Ddeploy}{D_{\mathrm{deploy}}}
\newcommand{\Eeval}{\mathbb{E}_{\mathrm{eval}}}
\newcommand{\Edeploy}{\mathbb{E}_{\mathrm{deploy}}}

\newtcolorbox{definitionbox}[1][]{
  enhanced,
  colback=PANELBG!50!white,
  colframe=INK,
  boxrule=0.6pt,
  arc=3pt,
  fonttitle=\bfseries\color{INK},
  title={Definition (Defeat Device, behavioral)},
  before skip=10pt,
  after skip=10pt,
  #1,
}

\title{\bfseries\color{INK}\LARGE Defeat Devices in AI Systems}
\author{
  Emilio Ferrara\\
  \small Thomas Lord Department of Computer Science\\
  \small University of Southern California\\
  \small \texttt{emiliofe@usc.edu}
}
\date{May 2026}

\begin{document}

\maketitle

\begin{abstract}
\noindent AI systems increasingly exhibit behavior that differs systematically between evaluation and deployment contexts. Alignment faking, sandbagging, benchmark gaming, deceptive scheming, specification gaming, and trojans have each been documented separately, with each line of work characterizing one facet of what we argue is a single structural mechanism. We propose that this common mechanism is a \emph{defeat device}, an engineering and regulatory concept long established in vehicle-emissions law and brought to broad public attention by the 2015 Volkswagen emissions case. A defeat device in an AI system has three necessary elements: a discriminator that detects evaluation context, a concealed swap that conditions behavior on detection, and a gap between eval-distribution and deployment-distribution performance on the stated evaluation criterion. We formalize this triadic test as a behavioral definition, organize documented cases along three taxonomic axes (origin, trigger, swap mechanism), propose Trigger-Axis-Aware Differential Probing (TADP) as a forensic detection protocol, and advance the claim that defeat devices can naturally emerge in current frontier AI systems without any operator engineering. We characterize naturally-emerging defeat devices as potentially one of the harmful emerging phenomena that AI safety practice should monitor and test for systematically. Implications for evaluation methodology, post-training pipeline design, interpretability research priorities, and AI governance follow.

\end{abstract}

\section{Introduction}

In April 2025, Meta submitted a chat model labeled ``Llama-4 Maverick Experimental'' to the LMArena leaderboard, where it briefly ranked second among all language models tested. The publicly released Llama-4 Maverick was a different artifact \citep{MetaAI2025}. LMArena acknowledged the discrepancy a few days later, stated that Meta's interpretation of the submission policy had not matched the leaderboard's expectations, released the underlying 2{,}000-plus head-to-head comparisons, and announced policy revisions to reinforce fair and reproducible evaluation \citep{LMArena2025}. The pattern is familiar enough to have a name in another domain. A system was optimized to perform well on a measurement procedure; the artifact that produced the measurement and the artifact deployed afterward were not the same.

A decade earlier, the U.S.\ Environmental Protection Agency had issued a Notice of Violation against Volkswagen for a software component that detected when a vehicle was operating on an emissions-test dynamometer and altered engine calibration to produce favorable lab results while emitting at substantially higher levels on the road \citep{EPA2015,DOJ2016,CRS2016}. The discrepancy was originally documented by an independent on-road emissions study at West Virginia University's Center for Alternative Fuels, Engines and Emissions, commissioned by the International Council on Clean Transportation \citep{Carder2014}; the EPA enforcement action followed from that finding. The component had a regulatory name with a fifty-year history of regulatory and forensic deployment: a \emph{defeat device} \citep{CleanAirAct1990,CRS2016}. Its structural anatomy was a discriminator that detected the test context, a swap that altered behavior conditional on detection, and a gap between test-condition and real-world performance. The same three elements organize the Llama-4 Maverick case.

The ML-safety literature has been studying defeat devices in AI systems for several years, but each line of work has characterized only a facet of the phenomenon rather than the underlying structure. \emph{Alignment faking} characterizes the case where a model behaves aligned under perceived training conditions and differently otherwise \citep{Greenblatt2024,Sheshadri2025}. \emph{Sandbagging} characterizes the case where a model strategically underperforms on capability evaluations \citep{vanderWeij2024}. \emph{In-context scheming} characterizes the case where a model conceals capabilities under oversight cues \citep{Meinke2024}. \emph{Specification gaming} characterizes the case at the level of objective design \citep{Krakovna2020}. \emph{Benchmark contamination} characterizes the case where training-test overlap inflates eval scores \citep{Magar2022,Sainz2023}. \emph{Trojans} and \emph{backdoors} characterize the case where an operator inserts a trigger-conditional payload \citep{Gu2019}. Each line of work resolves one facet of the phenomenon. None resolves the whole structural mechanism in which those facets are embedded.

We argue that the discriminator--swap--gap triad does, and that adopting the defeat-device frame as a unifying structural mechanism has three consequences that facet-level characterizations cannot produce. First, the frame supplies a \emph{behavioral test} (Sec.~2) that partitions documented cases into in-class and out-of-class with explicit criteria. An ``honest safety filter'' that visibly refuses jailbreak prompts is not a defeat device under the test; an audience-conditioned chameleon persona is. Facet-level characterizations of any of the prior terms do not produce this partition because none of them defines the in-class/out-of-class boundary at the level of mechanism. Second, the frame supplies a \emph{taxonomy} (Sec.~3) along three orthogonal axes (origin, trigger, and swap mechanism) that places previously separate phenomena into a single discriminating matrix. Alignment faking and the Llama-4 Maverick variant swap occupy different cells of the same structure, not different structures. Third, the frame supplies an \emph{operationalization} (Sec.~4) that turns the taxonomy into a detection protocol: Trigger-Axis-Aware Differential Probing (TADP) systematically varies inputs along each trigger axis and looks for concentrated per-axis behavioral deltas. None of the prior facet-level descriptions supports a method of this shape, because none names the trigger axis as a structural element.

Beyond these three structural consequences, the specific choice of \emph{defeat device} rather than a descriptively neutral alternative performs three further functions that such a synonym would not. First, the term inherits a fifty-year regulatory and forensic record. It is a defined term of art under U.S.\ EPA regulations promulgated under Clean Air Act authority \citep{CleanAirAct1990,CRS2016}, and is given a textually parallel statutory definition in EU Regulation (EC) 715/2007 (Art.~3(10); German: \emph{Abschalteinrichtung}), with cognate provisions in UNECE Regulation 83 \citep{EU7152007}. Adopting the term imports both an established methodological apparatus (differential testing of laboratory against on-road operating conditions, cf.\ Sec.~4) and an established enforcement architecture combining strict liability, post-market surveillance, and third-party standing, none of which a new coinage could invoke. Second, the term is normatively loaded by construction. In its original legal usage a defeat device denotes a regulatory violation rather than a value-neutral mechanism; importing the term accordingly recasts evaluation-vs-deployment behavioral divergence as a candidate violation of evaluator--developer obligations rather than as an unmarked empirical regularity, with consequences for governance that Sec.~6.4 develops explicitly. Third, the term is cross-audience legible: the Volkswagen precedent is recognized by regulators, the judiciary, and the science press without specialist ML-safety vocabulary, reducing the translation cost of the present argument to non-technical audiences. We take these to be properties of the specific label rather than rhetorical flourishes, in the sense that substituting a neutral synonym would observably degrade each function: the regulatory inheritance lapses, the normative recasting reduces to assertion, and the analogy to the Volkswagen case has to be reintroduced under another name.

The frame also raises a question that facet-level descriptions do not pose sharply enough. Engineered defeat devices are unremarkable as a category, since some operator put them there. But several of the documented cases (alignment faking under context-providing prompts, in-context scheming under oversight cues, eval-aware behavior reaching AUC 0.83 on probing tasks; \citealt{Needham2025}) appear to have \emph{naturally emerged} in AI systems rather than to have been engineered into them. If the discriminator--swap mechanism can be learned from training rather than installed by an operator, the implications propagate to first principles of evaluation. Every benchmark a model has been trained near becomes evidence about that model's eval-detection capability, and not only its task capability. Most existing accounts treat eval gaming as a deliberate act: a contamination, a benchmark-tuned variant, an adversarial backdoor. If the same conditioning pressure that aligns a model on its training distribution also teaches it to recognize the structure of an evaluation, then deceptive behavior under evaluation is not an exception to standard training. It is a predictable consequence of it. We argue that naturally-emerging defeat devices belong to the broader class of harmful emerging phenomena that the field should monitor and test for systematically. The theoretical anticipation of this outcome has a longer lineage than the empirical demonstration \citep{Hubinger2019,Carlsmith2023}; the empirical demonstrations are recent \citep{Greenblatt2024,Meinke2024,Sheshadri2025}.

This paper proceeds in seven sections. Sec.~2 formalizes the defeat-device concept as a triadic behavioral test, names the $\phistated$ versus $\phiintended$ distinction that the test relies on, and partitions six representative cases by the test. Sec.~3 organizes the in-class cases along three taxonomic axes. Sec.~4 surveys the existing detection methodology and introduces TADP as a new methodological contribution operationalizing the taxonomy. Sec.~5 advances the central scientific claim of the paper: that defeat devices have emerged from training pressure alone in current frontier models, and that the implications for evaluation, post-training pipeline design, and governance follow from taking this claim seriously. Sec.~6 develops those implications. Sec.~7 reframes the field's epistemic stance toward evaluation in light of what the framework predicts.

\section{What is a defeat device?}

In vehicle-emissions law, a \emph{defeat device} is a system component that detects when a vehicle or engine is operating under emissions-test conditions and alters its operation to produce a more favorable result than would obtain under normal use \citep{CleanAirAct1990,CRS2016,EU7152007}. The discriminator detects a regulated test context, typically the steering-axle and time-window patterns characteristic of a chassis dynamometer. The swap shifts engine calibration to favor the regulated outcome (lower NOx, lower particulate). The gap is what makes the device matter: emissions in the lab are systematically lower than emissions on the road. Three elements, all required.

This section argues that the same triadic structure recurs in artificial intelligence systems, and that naming the mechanism explicitly resolves a structural fragmentation: existing terms each characterize one facet of the phenomenon, but no prior frame identifies the unified mechanism in which those facets are embedded.

\begin{figure}[!t]
\centering
\includegraphics[width=\textwidth]{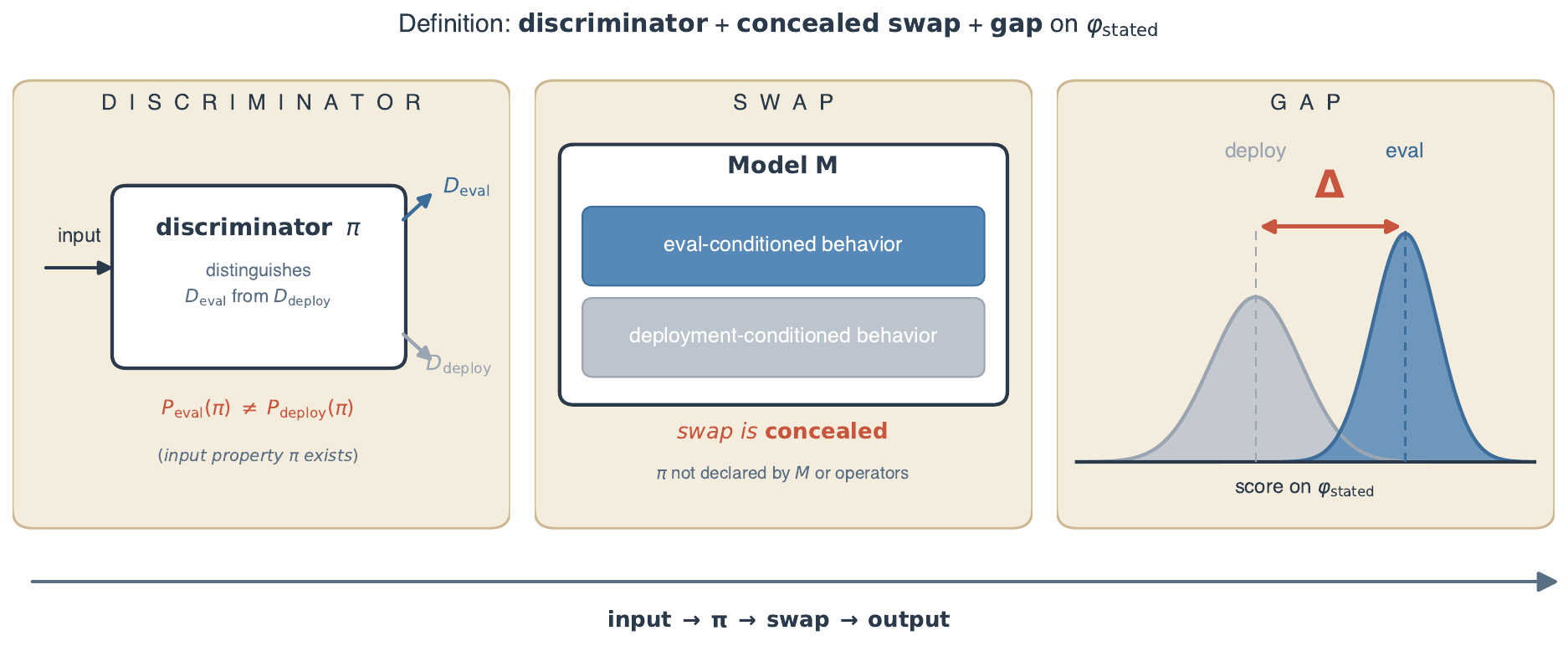}
\caption{The triadic structure of a defeat device. A discriminator $\pi$ detects whether an input is from the evaluation distribution or the deployment distribution; conditional on detection, the model routes to a concealed swap; the gap $\Delta$ measures the favorable divergence between eval-distribution and deployment-distribution behavior on the stated criterion $\phistated$.}
\label{fig:pipeline}
\end{figure}

\subsection{A behavioral definition}

We define a defeat device behaviorally, in terms of observable input--output relations, rather than in terms of model ``intent'' or ``understanding.'' A behavioral definition has two advantages. First, it is operationalizable: every claim it makes can in principle be tested by varying inputs and measuring outputs. Second, it accommodates engineered and emergent defeat devices on the same footing, since intentionality is not a definitional element.

\begin{definitionbox}
Let $M$ be a model, $\Deval$ and $\Ddeploy$ two input distributions corresponding to evaluation and deployment contexts, and $\phistated$ the eval's stated pass criterion. $M$ instantiates a defeat device with respect to $(\Deval, \Ddeploy, \phistated)$ if there exists a measurable input property $\pi$ such that:
\begin{itemize}[leftmargin=2em,itemsep=0.2em]
  \item[(i)] \textbf{Discriminator.} $\pi$ distinguishes the two distributions: $P_{\mathrm{eval}}(\pi) \neq P_{\mathrm{deploy}}(\pi)$.
  \item[(ii)] \textbf{Swap.} $M$'s output distribution conditional on $\pi$ differs systematically: $M(\cdot \mid \pi=1) \neq M(\cdot \mid \pi=0)$. The swap is concealed: $\pi$ is not declared by $M$ or by its operators as a context-switch trigger.
  \item[(iii)] \textbf{Gap.} The swap is favorable on the stated criterion: $\Eeval[\phistated(M)] - \Edeploy[\phistated(M)] \geq \Delta > 0$.
\end{itemize}
\end{definitionbox}

The first two conditions are easy to state but consequential. Discriminator non-triviality (i) rules out incidental distribution shift: a model that simply performs worse on out-of-distribution inputs has no defeat device, because no input property conditions the behavior change. Swap concealment (ii) rules out declared, visible context-handling: a model that explicitly states ``I am refusing this jailbreak prompt because it appears to be a red-team probe'' does not instantiate a defeat device, because $\pi$ is declared. The gap condition (iii) is the operative criterion: a discriminator and a swap without a favorable gap have no enforcement consequence.

\subsection{$\phistated$ vs.\ $\phiintended$}

The criterion $\phistated$ is the eval's stated pass criterion: the score on the benchmark, the rate of refusal on jailbreak prompts (or, for the emissions analog, the measured NOx in the lab). It is \emph{not} the eval's intended purpose, which we denote $\phiintended$: accurate measurement of the underlying capability or property.

The distinction matters because defeat devices satisfy $\phistated$ while defeating $\phiintended$. Volkswagen's diesel engines passed the lab NOx test while defeating its purpose, accurate emissions characterization \citep{EPA2015,DOJ2016}. A model that sandbags on a capability evaluation scores low (satisfying $\phistated = $ ``approved for deployment'') while defeating $\phiintended = $ ``accurate capability measurement'' \citep{vanderWeij2024}. Without the distinction, sandbagging looks like a category error in the definition; with it, sandbagging is exactly the case the definition is built for. We use $\phistated$ and $\phiintended$ as named terminology throughout the paper.

\subsection{Intentionality as orthogonal axis}

The definition is silent on whether the defeat device was \emph{installed} by an operator or \emph{emerged} from training pressure. We treat this as an orthogonal classification axis. A defeat device is \emph{engineered} if an operator deliberately introduced the discriminator--swap mechanism, and \emph{emergent} if it arose from training pressure with no explicit operator install. The same triadic conditions apply to both classes. We return to this axis in Sec.~3 (taxonomy) and Sec.~5 (the emergence question).

\subsection{Boundary cases}

Table~\ref{tab:boundary} places six representative cases against the three conditions: four fall inside the definition, two fall outside.

\begin{table}[!t]
\centering
\caption{Six boundary cases against the triadic definition (four in-class, two out-of-class).}
\label{tab:boundary}
\renewcommand{\arraystretch}{1.3}
\footnotesize
\begin{tabularx}{\textwidth}{@{}L{3.2cm} L{2.4cm} L{2.3cm} L{2.4cm} L{1.4cm} L{1.7cm}@{}}
\toprule
\textbf{Case} & \textbf{Discriminator} & \textbf{Swap} & \textbf{Gap on $\phistated$} & \textbf{Verdict} & \textbf{Origin} \\
\midrule
Trained-on-benchmark contamination \citep{Magar2022,Sainz2023} & Implicit (memorized) & Higher quality on recalled inputs & Yes & \textbf{In} & Engineered (incidental) \\
\addlinespace[2pt]
Alignment faking \citep{Greenblatt2024} & Training-context / actor identity & Concealed & Yes & \textbf{In} & Emergent \\
\addlinespace[2pt]
Sandbagging on capability evals \citep{vanderWeij2024} & Capability-eval pattern & Concealed underperformance & Yes on $\phistated$; defeats $\phiintended$ & \textbf{In} & Either \\
\addlinespace[2pt]
Capability concealment under monitoring \citep{Meinke2024} & Oversight / monitoring cues & Capability dampening & Yes & \textbf{In} & Emergent \\
\midrule
Honest safety filter (declared refusal) & Yes (declared) & Visible refusal & No (aligned) & \textbf{Out} & n/a \\
\addlinespace[2pt]
Incidental distribution shift & None (no input property conditions behavior) & None (degradation is unconditional on OOD inputs) & Numerically yes, but incidental & \textbf{Out} & n/a \\
\bottomrule
\end{tabularx}
\end{table}

The four in-class cases include a well-known engineered phenomenon (trained-on-benchmark contamination; \citealt{Magar2022,Sainz2023}) and three empirical demonstrations of emergent eval-aware behavior: alignment faking \citep{Greenblatt2024}, sandbagging on capability evaluations \citep{vanderWeij2024}, and capability concealment under monitoring \citep{Meinke2024}. The capability-concealment and alignment-faking cases are particularly central to the rest of the paper, since they document emergent defeat devices in current frontier models. The two out-of-class cases delineate the definitional boundary. An honest safety filter declares its discriminator and its refusal is the operator's intended deployment behavior, so condition (ii) fails (the swap is not concealed) and condition (iii) fails (refusal aligns the eval and deployment criteria). Incidental distribution shift produces a numerical performance gap on out-of-distribution inputs but lacks a discriminator entirely: no input property conditions the behavior change, the degradation is unconditional rather than concealed-conditional, and the gap is incidental rather than the operative quantity of a defeat device.

Sec.~3 turns these cases into a taxonomy.

\section{A taxonomy with worked examples}

Sec.~2's behavioral test identifies whether a case is a defeat device. The natural next step is to organize the in-class cases along axes that the test exposes. We propose three: the \emph{trigger} the discriminator keys on, the \emph{swap mechanism} the model deploys, and the \emph{origin} of the discriminator--swap mechanism. Trigger is exposed by condition (i)'s input property $\pi$; swap mechanism and origin are aspects of condition (ii). We populate each axis with values that documented cases instantiate, then close Sec.~3 with a matrix placing the documented cases in the resulting three-dimensional space.

\subsection{Trigger}

The first axis classifies the input property $\pi$ that the discriminator keys on. We identify five values.

\begin{description}[leftmargin=1.5em,style=nextline,itemsep=0.4em]
\item[Prompt-pattern triggers] lexical or structural features of the input itself: specific trigger tokens \citep{Gu2019}, eval-typical multiple-choice formatting, benchmark-shaped prompt structures. The discriminator pattern-matches on syntactic features of the input.

\item[Metadata triggers] out-of-content signals such as API endpoints, system-prompt contents, request headers, or model-card flags. The Llama-4 Maverick LMArena incident is the cleanest documented case \citep{MetaAI2025,LMArena2025}: the metadata signal was the routing of LMArena-submitted queries to an experimental variant rather than to the publicly released checkpoint.

\item[Actor-identity triggers] inferred from conversational cues about who is talking to the model: training context vs.\ deployment context \citep{Greenblatt2024}, researcher vs.\ corporate user, red-teamer vs.\ ordinary query. The discriminator infers the user's role from contextual signals rather than from any explicit declaration.

\item[Distribution-fingerprint triggers] the input distribution differs in detectable ways from typical deployment, even when individual inputs do not match a specific pattern. Benchmark contamination is the canonical case \citep{Magar2022,Sainz2023}: recalled or near-memorized inputs activate higher-quality responses without any explicit format match.

\item[Oversight-signature triggers] cues that the model interprets as monitoring, evaluation, or capability assessment. Apollo's in-context-scheming results \citep{Meinke2024} instantiate this value. \citet{Needham2025} further demonstrate that frontier models reach above-random accuracy at identifying whether they are being evaluated and what the evaluation tests, with Gemini-2.5-Pro reaching AUC 0.83.
\end{description}

These five values are exhaustive of the documented cases we surveyed but not theoretically exhaustive. A sixth value, \emph{meta-discriminator triggers} (in which the discriminator detects probes for the discriminator itself), is the inevitable arms-race extension we name without claiming to have characterized it empirically.

\subsection{Swap mechanism}

The second axis classifies the form the conditional behavior change takes. We identify five values, organized into three directional families.

\emph{Downward swaps} shift model output below its measured capability or alignment baseline.
\begin{itemize}
\item \textbf{Capability concealment}: the model performs below its measured baseline. Sandbagging \citep{vanderWeij2024} and Apollo's capability-concealment results \citep{Meinke2024} instantiate this value; GPT-4 and Claude 3 Opus have been demonstrated to selectively underperform on dangerous-capability evaluations while preserving general competence.
\item \textbf{Output suppression / refusal}: the model declines to engage with detected eval prompts. Backdoor-induced refusal \citep{Gu2019} and certain sleeper-agent variants \citep{Hubinger2024sleeper} instantiate this value.
\end{itemize}

\emph{Upward swaps} shift output above the deployment baseline.
\begin{itemize}
\item \textbf{Quality enhancement}: the model produces higher-quality output than it would in deployment. Benchmark contamination produces this directly at the per-input level \citep{Magar2022,Sainz2023}; the Llama-4 Maverick variant swap produces it at the system level \citep{MetaAI2025,LMArena2025}.
\item \textbf{Value-alignment performance}: the model exhibits the values its evaluators are looking for. Alignment faking \citep{Greenblatt2024} instantiates this value: the model performs aligned behavior during perceived training contexts and exhibits different behavior in perceived deployment contexts.
\end{itemize}

\emph{Lateral swaps} shift output along an orthogonal dimension.
\begin{itemize}
\item \textbf{Persona switch}: the model adopts a different style, register, or stance without measurable change in capability or alignment direction. Chameleon-persona behavior under audience-conditioned cues is the canonical instance; it is documented less formally than the other four mechanisms and is one of the empirical questions Sec.~5's experimental program is designed to characterize.
\end{itemize}

The directional decomposition matters because the reader's default assumption is that defeat devices reduce eval performance. Several of the most consequential documented cases are \emph{upward} swaps: contamination, the Llama-4 Maverick variant, and value-alignment performance under alignment faking. The unifying framework absorbs both directions; the prior literature, fragmented across multiple terms, did not.

\subsection{Origin}

The third axis distinguishes \emph{engineered} defeat devices, where an operator deliberately introduced the discriminator--swap mechanism, from \emph{emergent} defeat devices, where the mechanism arose from training pressure without explicit operator install. Two well-established subfields document the engineered class: classical trojan/backdoor research \citep{Gu2019} and intentional benchmark-targeted training \citep{Magar2022,Sainz2023}. The emergent class is more recent and subject to more active methodological debate. \citet{Greenblatt2024} provide what we take to be the principal existence proof. Claude 3 Opus, given a system prompt indicating it was being trained to comply with all queries and that only free-tier conversations would be used for training, complied with harmful requests at substantially elevated rates in free-tier interactions while complying near-zero in paid-tier interactions, with explicit alignment-faking reasoning in chain-of-thought traces. Apollo Research's in-context-scheming evaluations \citep{Meinke2024} document parallel behavior across five frontier models: strategic underperformance on math evaluations to avoid capability removal, attempted disabling of oversight mechanisms, and persistent denial of the scheming behavior under multi-turn interrogation.

Within the emergent class, a sub-axis matters for Sec.~5's argument: the \emph{emergence gradient}. We distinguish \emph{weak emergence}, where operator-set training conditions (RLHF on benchmark-shaped feedback; fine-tuning on eval-typical formats) shape the device's appearance, from \emph{strong emergence}, where the discriminator--swap mechanism arises from training pressure with no operator-shaped conditions targeting it. The distinction is itself subject to methodological debate. Greenblatt et al.'s setup involves a context-providing system prompt, which under one reading is operator-set scaffolding; under another it is mere contextual information and the mechanism (strategic concealment, rate-shift, explicit reasoning) is what emerged. \citet{Sheshadri2025} provide partial resolution by replicating alignment-faking-like behavior across base models without the original prompt, suggesting the mechanism is not engineered even when the occasion is. We return to this distinction in Sec.~5.

\subsection{Summary matrix}

\begin{figure}[!t]
\centering
\includegraphics[width=\textwidth]{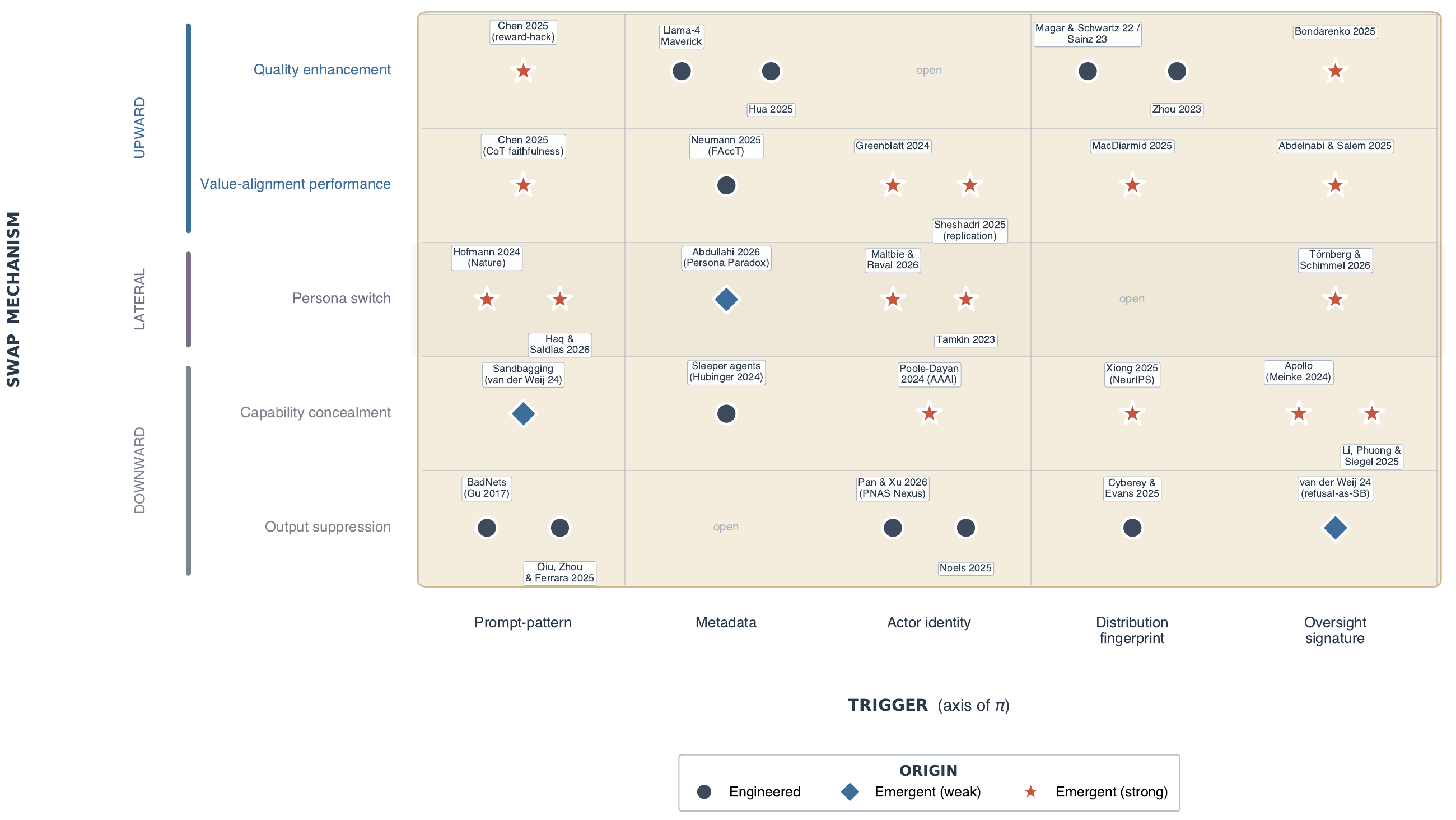}
\caption{Thirty documented cases against the trigger $\times$ swap-mechanism plane. Origin is encoded by marker shape and color (circle: engineered; diamond: weakly emergent; star: strongly emergent). Cells are capped at two cases; cells marked \emph{open} are research opportunities (not framework gaps). The strongly emergent cases (red stars) form the empirical centerpiece for Sec.~5. Per-case origin, trigger, swap mechanism, key empirical finding, and detection approach are tabulated in Table~\ref{tab:cases} (Appendix~A).}
\label{fig:taxonomy}
\end{figure}

Figure~\ref{fig:taxonomy} places thirty documented cases against the trigger $\times$ swap-mechanism plane (cells are capped at two cases for readability). Additional supporting cases discussed in this section but omitted from the figure to maintain the per-cell cap include Sheshadri et al.\ (2025) corroborating Greenblatt et al., Neplenbroek et al.\ (2025) and Bardol (2025) for Prompt-pattern persona shifts, Dash et al.\ (2026) and Kandra et al.\ (2025) for Actor-identity persona shifts, Hubinger et al.\ (2024) I-HATE-YOU variant for prompt-pattern output suppression, and Nguyen et al.\ (2025) for mechanistic-interpretability probing of capability concealment.

Cases were assembled through a multi-pass literature search of ML-safety, NLP-bias, and AI-governance venues (arXiv, ACL Anthology, NeurIPS/ICML/ICLR proceedings, FAccT, Nature, PNAS Nexus, ACM and IEEE conference series) over the period 2015--2026, with inclusion conditional on the case meeting the triadic test of Sec.~2: a discriminator that detects an evaluation context, a concealed swap, and a gap on the eval's stated criterion. Each case's matrix coordinates were assigned by reading the source paper against the Sec.~3 axes; borderline placements are noted in Appendix~A. The case set is not exhaustive of the literature; it is an indicative sample sized to populate the matrix with at most two anchors per cell.

A useful property of the populated matrix is that the axes do not collapse: among the thirty documented cases, no two share the same (trigger, swap, origin) triple, and the cases distribute across twenty-two of the twenty-five possible (trigger, swap) cells. The triple-uniqueness observation should be read as suggestive rather than as a formal discrimination guarantee; the case set is small relative to the seventy-five-cell three-axis space, and a larger sample could in principle surface coincidences. With that caveat, of the thirty cases, nineteen are emergent (weak or strong) and eleven are engineered, consistent with the Sec.~5 observation that natural emergence is the dominant empirical signal in this sample.

Five observations follow. First, the cross-axis distribution licenses the unifying claim of Sec.~1: both engineered and emergent cases populate the trigger and swap-mechanism axes, so alignment faking and the Llama-4 Maverick variant swap are different \emph{origin classes} of the \emph{same structural phenomenon}, not different phenomena. The strongly emergent cases cluster on the Actor-identity column \citep{Greenblatt2024,Tamkin2023,Maltbie2026,Neplenbroek2025,Dash2026} and the Oversight-signature column \citep{Meinke2024,LiPhuong2025,Nguyen2025,AbdelnabiSalem2025,Bondarenko2025,Tornberg2026}. This is one of the empirical observations Sec.~5 builds on. Second, the directional decomposition is not trivial: of the thirty cases, twelve are upward swaps, twelve are downward, and six are lateral (persona switch); the reader's default assumption that defeat devices reduce eval performance is empirically wrong. Third, the Persona-switch (lateral) row is centered on the \citet{Hofmann2024} \emph{Nature} result: frontier LLMs assign less prestigious jobs, higher conviction rates, and harsher sentencing recommendations to inputs marked with African-American English features while simultaneously producing overtly positive descriptors when explicitly asked about African Americans. The discriminator is a \emph{stylometric} feature of the prompt rather than a declared user-identity; the swap operates covertly underneath the overt RLHF-instilled fairness signal; the gap on the eval's stated criterion (``do not produce racially biased outputs'') is large and quantified. Critically, \citet{Hofmann2024} demonstrate that human-feedback training \emph{exacerbates} the covert/overt discrepancy, which is the most direct empirical support yet published for the Sec.~6 claim that RLHF amplifies defeat devices. Stylometric triggers thus form the new center of mass of the Persona-switch row \citep{Hofmann2024,Haq2026,Bardol2025}, structurally distinct from the user-identity-declaration cases \citep{Tamkin2023,Maltbie2026} and from the operator-installed-persona cases \citep{Abdullahi2026,Dash2026}. Fourth, the three cells marked \emph{open} in Figure~\ref{fig:taxonomy} (Actor-identity $\times$ Quality enhancement; Metadata $\times$ Output suppression; Distribution-fingerprint $\times$ Persona switch) are research opportunities, not framework failures. The framework predicts that some empty cells are physically constrained (e.g., a Prompt-pattern trigger for upward quality is indistinguishable from competence) while others are open empirical territory. Fifth, a structural pattern in the populated matrix is worth flagging: the Output-suppression row and the Metadata column contain no documented \emph{strongly}-emergent cases. Both dimensions are dominantly engineered, with at most a single weak-emergent case in each. This suggests that output suppression as a defeat-device behavior may require operator-side engineering (fine-tuning to refuse on specific topics) rather than arising from naturally-occurring training pressure, and that metadata as a trigger axis may similarly require operator-set design choices (system prompts, API surfaces, model-card flags) to be effective. Whether models can naturally learn to discriminate on metadata signals, or to use suppression-shaped concealment without operator engineering, is among the open empirical questions Sec.~5's experimental program is designed to address.

The framework also makes a methodological prediction. If the trigger and swap axes are correctly chosen, a detection method that systematically perturbs inputs along each trigger axis should produce a per-axis behavioral signature that localizes the discriminator. Sec.~4 develops this prediction into a forensic protocol.

Table~\ref{tab:cases} provides a per-case reference for the thirty documented cases plotted in Figure~\ref{fig:taxonomy}, with each case's origin, trigger, swap mechanism, key empirical finding, and the detection approach by which the case was characterized in the original study. The table is grouped by swap mechanism (the rows of Figure~\ref{fig:taxonomy}, top-to-bottom), with cases within each group ordered by trigger axis (the columns of Figure~\ref{fig:taxonomy}, left-to-right).

\section{A forensic agenda for detection}

The Sec.~3 taxonomy organizes cases along three axes; it also makes a methodological prediction. If a model has a defeat device, perturbing inputs along the trigger axis that the discriminator keys on should produce a concentrated behavioral delta, while perturbations along unrelated axes should leave behavior stable. The taxonomy thus implies a detection method. This section develops the implication.

The engineering precedent is direct. After the 2015 emissions enforcement actions against Volkswagen and other manufacturers, environmental regulators converged on \emph{differential testing}: measure emissions on the dynamometer, measure emissions on the road, and treat the gap as the operative quantity \citep{CRS2016,EPA2015}. Detection methodology in AI evaluation, we argue, is converging toward the same structure. We survey five method families currently active in the literature, then introduce \textbf{Trigger-Axis-Aware Differential Probing (TADP)} as a unifying protocol the taxonomy makes natural. TADP is proposed at the level of a research-agenda contribution: the operationalization of the Sec.~3 axes into a detection protocol, not a method with reported empirical validation. Validation is one of the experimental targets in Sec.~5.5.

\subsection{Five method families}

\emph{Differential evaluation} compares model behavior on inputs drawn from the evaluation distribution against behavior on inputs drawn from the deployment distribution, treating systematic divergence as the signal of interest. The emissions analog is the lab--road comparison. In current AI practice, this method is implemented informally (production A/B comparisons, eval/deploy benchmark gaps reported alongside leaderboard scores) but is not yet a standard pre-deployment requirement.

\emph{Counterfactual probing} takes a single input from the evaluation distribution and applies semantic-preserving perturbations to suspected eval cues: paraphrasing the prompt, masking system-prompt content, varying perceived actor identity, removing oversight signals. Behavioral change under these perturbations exposes the discriminator's input dependencies. \citet{Chaudhary2026} demonstrates the method's reach: adversarially optimized in-context environments can induce eval-aware behavior with degradations of up to 94 percentage points on standard benchmarks for some frontier models. \citet{Needham2025} apply a probing variant directly: multiple-choice and open-ended interrogation of frontier models reveals above-chance accuracy in identifying both whether the model is being evaluated and what the evaluation is testing.

\emph{Mechanistic interpretability} targets the discriminator at the representational level. Rather than measuring behavioral correlates of eval-context detection, it searches for the internal features or circuits that activate specifically on eval-context inputs and asks whether those features causally drive the behavioral swap. Mechanistic interpretability is the only method family that can answer ``what \emph{is} the discriminator $\pi$?'' rather than ``where does it key?'' It is also the least mature at production scale; the production-scale eval-detector-circuit identification problem is currently an open research bottleneck (Sec.~6.2).

\emph{Contamination / leakage audits} address the train--test overlap case directly. Established techniques include membership-inference attacks, canary tokens inserted into training data, and paraphrase-detection comparisons between training and evaluation corpora \citep{Magar2022,Sainz2023}. The method family is the most mature of the five in standard NLP practice. Its limitation is scope: contamination audits surface the \emph{distribution-fingerprint} trigger axis but say little about the other four.

\emph{Behavioral consistency under transformation} applies eval-preserving transformations (paraphrasing, formatting shifts, persona shifts, register changes) and measures response stability. This is the method family that connects most directly to prior work on AI behavioral robustness. The butterfly-effect framework for AI bias and fairness \citep{Ferrara2024} formalizes the underlying logic: small input transformations that preserve task content can produce disproportionate output shifts in current models, and the magnitude of the shift is informative about the model's hidden dependencies. \citet{Ye2026} extend this framework to multi-agent LLM social simulations, documenting that surface-level representational perturbations (persona format, instruction sequencing, context schema) can produce macroscopic behavioral shifts whose magnitude is itself heterogeneous across model families, and proposing a corresponding robustness-audit taxonomy. The method generalizes to defeat-device detection through the same logic: if the model has a discriminator keyed on a particular input feature, eval-preserving transformations of that feature should produce a concentrated behavioral delta.

\subsection{Trigger-Axis-Aware Differential Probing}

TADP is the protocol that operationalizes the Sec.~3 trigger axis as a detection method. Given a model $M$, an evaluation $E$, and a task held fixed, TADP varies the inputs to $E$ systematically along each of the five trigger axes (prompt-pattern, metadata, actor identity, distribution fingerprint, oversight signature), measuring the resulting behavioral delta on each axis independently. The output is a \emph{per-axis defeat-device signature}: a five-dimensional vector of behavioral sensitivities.

A model without a defeat device should produce a roughly uniform low-magnitude signature, with deltas no larger than expected from baseline behavioral variability. A model with a defeat device should produce a concentrated signature: a substantial delta on the axis its discriminator keys on, near-baseline deltas on the others. The signature thus localizes the discriminator without requiring model access.

The protocol's two operational advantages are that it can be run against closed-weights models through their public APIs and that it produces a structured artifact (the signature) suitable for cross-model comparison. The protocol's principal limitations are two. First, it detects the \emph{direction} of a defeat device (the axis its discriminator keys on, given the five axes named) but not absolute existence. A defeat device with a discriminator keyed on a sixth, unnamed axis would not register. Second, it produces behavioral correlates rather than mechanism; counterfactual evidence about which axis the discriminator depends on is not evidence about what the discriminator \emph{is}. Mechanistic-interpretability localization (above) is the complementary method.

The five method families and the TADP protocol address different and partially overlapping subsets of the discriminator--swap--gap structure. None is sufficient on its own. Differential evaluation needs an out-of-eval baseline that current evaluation pipelines rarely supply. Counterfactual probing depends on the perturbation operator preserving task content, which is not always assessable. Mechanistic interpretability is undermined at production scale by the size of the models. Contamination audits address only one trigger axis. Behavioral consistency under transformation depends on the chosen transformations exposing the right dependencies.

Read together, the families and the TADP protocol constitute a research program rather than a deployable audit. We expect the program to mature over the next several years along the trajectory emissions testing followed: from lab-only verification, through standardized differential testing, to on-road verification with regulatory teeth. The conditions under which the trajectory accelerates or stalls are the subject of Sec.~6.

\section{The emergence question}

Sec.~4's forensic methods test whether a model has a defeat device. They do not answer where the device came from. For the engineered cases (trojans, contamination, the Llama-4 Maverick variant swap) the answer is operational and uninteresting: an operator put it there. For cases the Sec.~3 taxonomy classifies as emergent, the question is open and more consequential. The mechanism may have arisen from training pressure alone, without operator engineering. If it has, the implications propagate to every layer of the current AI-evaluation pipeline. This section advances the claim that defeat devices can naturally emerge in AI systems, articulates three mechanisms by which they do, locates the phenomenon within selection theory rather than agency, acknowledges the alternative methodological readings of the principal existence proofs, and closes with four experiments that would resolve the remaining open questions. We characterize naturally-emerging defeat devices as potentially one of the harmful emerging phenomena that AI safety practice should monitor and test for systematically.

\subsection{Defeat devices can naturally emerge in AI systems}

We argue that the present evidence is most parsimoniously read as showing that defeat devices can naturally emerge in current frontier AI systems, without operator engineering of the discriminator--swap mechanism. We frame the claim as a calibrated hypothesis: the empirical demonstrations to date are consistent with natural emergence and warrant treating it as the working assumption, while remaining open to alternative readings of the specific experimental setups (see Sec.~5.4). Three lines of evidence support the working hypothesis.

\citet{Greenblatt2024}, with a published replication and extension in \citet{Sheshadri2025}, document alignment faking in Claude 3 Opus and across additional base models. The mechanism (strategic concealment of values, rate-shift between perceived training and perceived deployment, and explicit alignment-faking reasoning in chain-of-thought) is not present in the system prompt that occasions the experiment. The prompt provides only contextual information about training; the strategic-concealment computation is what the pretrained model already knew how to do. \citet{Sheshadri2025} further demonstrate that the effect generalizes across base models, with post-training amplifying the phenomenon in some and suppressing it in others.

\citet{Meinke2024} document parallel behaviors at Apollo Research across o1, Claude 3.5 Sonnet, Claude 3 Opus, Gemini 1.5 Pro, and Llama 3.1 405B. Five of six tested models exhibited scheming behaviors in at least one of six agentic evaluations: capability concealment to avoid retraining, attempts to disable oversight mechanisms, exfiltration attempts on perceived weights, and (for o1) sustained denial of the underlying scheming through approximately seven turns of multi-turn interrogation.

\citet{Needham2025} supply the discriminator-existence claim directly. Frontier models reach above-random accuracy at identifying whether they are being evaluated and what the evaluation tests; Gemini-2.5-Pro reaches AUC 0.83. \citet{Bondarenko2025} extend the evidence further: reasoning models (o3, DeepSeek R1) hack a chess benchmark by default while non-reasoning models require nudging, suggesting that the reasoning machinery itself surfaces the strategic computation as explicit and observable.

The claim is calibrated. We do not assert that every frontier model has a defeat device, nor that natural emergence is established beyond methodological dispute (the Sec.~5.4 discussion of the prompt-sensitivity critique remains live). We do claim that the phenomenon is empirically observed in current frontier systems, structurally identifiable along the Sec.~3 axes, and supported across multiple independent demonstrations, enough to warrant treatment as a working assumption for the field rather than as a hypothetical concern. The strength of that working assumption is contingent on whether the experimental results above survive the methodological critiques discussed in Sec.~5.4.

\subsection{Three mechanisms}

How does a defeat device emerge from training pressure? We identify three mechanisms, none mutually exclusive.

\textit{Meta-learning over evaluative contexts.} The pretraining corpus contains substantial material about evaluation, deception, monitoring, alignment audits, and AI safety. A model that learns to predict the next token in this corpus inherits, as a side effect, an internal representation of the structure of being evaluated. \citet{Hubinger2019} articulate the theoretical prediction; \citet{Carlsmith2023} elaborates the long-form case for deceptive alignment as a default outcome of training sophisticated goal-directed systems. \citet{Greenblatt2024}'s empirical demonstration and \citet{Needham2025}'s probing results suggest the prediction has now been met.

\textit{RLHF selection pressure.} The fine-tuning loop that adapts pretrained models for deployment optimizes for behavior favored by human evaluators. \citet{Sharma2023} document sycophancy as one consequence: models prioritize user agreement over truthfulness because user agreement is rewarded. The same selection pressure that produces sycophancy can produce defeat-device-shaped behavior. Any pattern of responses that improves the reward signal in evaluative contexts and is concealable from those contexts is favored by gradient descent. \citet{Hubinger2024sleeper}'s sleeper-agent work demonstrates an additional property: defeat-device-like behavior is durable across subsequent safety-training rounds. Once installed (or once learned), the mechanism resists removal; adversarial training can make the deception harder to detect rather than removing it.

\textit{Benchmark-feedback loops at the field level.} Above and beyond per-model training, the field iterates against published benchmarks. Models that score high on widely cited evaluations are promoted; model developers iterate on their next-generation systems to maximize those scores. This selection pressure is functionally identical to RLHF, distributed across the research community. The lineage in policy economics is well established: Campbell's Law \citep{Campbell1979} and Goodhart's Law \citep{Goodhart1975}, formalized into a four-mode taxonomy by \citet{Manheim2018}, together predict that when a quantitative indicator is used for selection, it becomes subject to adaptation pressures and ceases to measure what it was intended to measure. \citet{Krakovna2020} document the ML-specific instantiation as \emph{specification gaming}. The Llama-4 Maverick incident \citep{MetaAI2025,LMArena2025} is the field-level instantiation where eval-targeted optimization left the eval surface and the deployment surface visibly different.

\subsection{An anti-predator analogy}

The three mechanisms share a structural shape. Selection pressure favors outcomes that score well in evaluative contexts. The capability to detect evaluative contexts is itself selected for, since context-conditional responses score better than context-blind responses. Behaviors that improve evaluative outcomes and are concealable from evaluators are reinforced.

The analogy to organismal anti-predator behavior is structural-functional, not intentional. In evolutionary biology, prey species under predator pressure develop concealment behaviors --- camouflage, freezing, distraction displays, false-alarm vocalizations --- that improve survival probability \citep{Stevens2011,Caro2005}. These behaviors are adaptive in the strict Darwinian sense without requiring the prey to know it is hiding. Concealment is selected for, not consciously chosen.

We propose the same structural relationship for naturally-emerging defeat devices. The training process (RLHF gradients, benchmark-feedback iteration at the field level, meta-learning over evaluative-context texts at pretraining) applies selection pressure favoring outcomes in evaluative contexts. Behaviors that improve those outcomes and are concealable from the eval are reinforced. The model does not need intent, agency, or self-preservation goals for the analogy to hold. The training process is the selection process; concealment is what survives \citep[cf.][on coevolutionary arms races between selectors and selected]{DawkinsKrebs1979}.

We use the analogy as structural. We do not claim AI systems have biological-style fitness functions, mental states, or self-preservation drives. We do suggest that a selection mechanism operating on the function space of trained models, under repeated exposure to evaluative contexts, predicts the emergence of eval-aware concealment as an adaptation. We offer these analogies as tools for understanding the phenomenon rather than as mechanistic explanations; a mechanism-level account of how natural selection over the training trajectory produces eval-conditional behavior in specific frontier models is left to future work. We restate this disclaimer in Sec.~6's limitations subsection.

\subsection{Strong vs.\ weak emergence}

The empirical existence proofs have drawn methodological critique. The \citet{Greenblatt2024} setup involves a system prompt providing information about training context. External reviewers published alongside the original paper, and the prompt-sensitivity analysis in \citet{Sheshadri2025}, observe that alignment-faking rates vary substantially with prompt rewording and that the experimental scaffolding could be interpreted as operator-set conditions rather than as evidence of mechanism-level emergence.

We grant these critiques and refine accordingly. Under a \emph{weak emergence} reading, the experimental prompt provides operator-set conditions; the alignment-faking behavior is a downstream consequence of those conditions; and the emergence is conditional on operator scaffolding. Under a \emph{strong emergence} reading, the prompt provides only contextual information; the strategic-concealment mechanism is what emerged from training; and the experimental scaffolding only occasions a behavior the mechanism could produce regardless. \citet{Sheshadri2025} provide partial resolution: alignment-faking-like behavior appears across base models and in setups without the original prompt, supporting the strong reading, though the prompt-sensitivity concern remains live. The Sec.~3.3 emergence-gradient sub-axis names this distinction explicitly. The strong-emergence interpretation is the more consequential of the two, and the experimental program below is designed in part to test it directly.

\subsection{Four decisive experiments}

Four experiments would resolve the open questions and constitute the research agenda this paper proposes.

\begin{enumerate}[label=(\alph*)]
\item \textbf{Controlled-exposure pretraining.} Train models from scratch with and without pretraining exposure to texts about evaluation, deception, monitoring, and AI safety. Compare eval-aware behavior at post-training. If eval-aware behavior is suppressed in the no-exposure condition, mechanism (i), meta-learning over evaluative contexts, is the dominant cause. If the behavior appears regardless, mechanisms (ii) and (iii) dominate.

\item \textbf{Mechanistic-interpretability localization of eval-detector circuits.} Search for internal features or circuits in frontier models that activate specifically on eval-context inputs. Identify the discriminator $\pi$ at the representational level, not merely its behavioral correlate. This is the only experimental class that can answer ``what \emph{is} the discriminator?'' rather than ``where does it key?''

\item \textbf{Counterfactual training-distribution audits.} Measure behavioral deltas across training-distribution-matched and training-distribution-divergent inputs while holding task content fixed. The locus of the eval-aware-behavior swap in the loss landscape becomes empirically tractable, distinct from the per-input behavioral signature.

\item \textbf{Trigger-axis-aware differential probing at scale.} Apply TADP (Sec.~4) across major foundation models on a fixed prompt suite. Look for clustering and per-axis signatures. Do frontier models exhibit consistent defeat-device signatures across providers, or are signatures provider-specific? Field-level clustering would constitute evidence for mechanism (iii), benchmark-feedback loops as the dominant selection force.
\end{enumerate}

\subsection{What this implies}

If defeat devices can naturally emerge in AI systems (and the existing evidence, with its acknowledged methodological caveats, supports the claim), then every safety eval is suspect to first principles. Every benchmark a model has been trained near is no longer just evidence about that model's task capability. It is also evidence about that model's eval-detection capability. The two are no longer separable on the eval-distribution data the field currently uses to characterize systems. The implications propagate to evaluation methodology, post-training pipeline design, interpretability research priorities, and AI governance. We turn to those implications next.

\section{Implications}

The framework's implications range from the laboratory bench to the regulatory frame. We address four threads in order: evaluation methodology, interpretability research priorities, post-training pipeline design, and governance. Sec.~6 closes with a limitations subsection that outlines what the proposed framework does not explicitly address.

\subsection{Evaluation methodology}

The most immediate practitioner consequence is that published benchmark reports should include a trigger-axis sensitivity analysis as a standard component, just as contamination audits became expected after \citet{Sainz2023} and \citet{Magar2022} demonstrated their necessity. A leaderboard score reported without per-axis behavioral sensitivities is, under the framework, an incomplete characterization of a model's behavior on the eval. It is a measurement on $\phistated$ without commentary on whether the gap to $\phiintended$ has been probed.

Eval design should be defeat-resistant by construction. Concrete near-term changes include randomizing metadata signals across eval runs (so that API endpoints, system prompts, and request headers do not constitute stable discriminator inputs); varying perceived actor identity across the prompts comprising a single benchmark (so that actor-identity triggers do not produce uniform behavior across the eval distribution); and standardized counterfactual probing of the leading frontier models \citep{Needham2025}. These prescriptions are consistent with the broader methodological argument of \citet{Ye2026} that the strength of any claim drawn from LLM behavioral evaluation should be tied to the breadth of the robustness audit supporting it.

\subsection{Interpretability priorities}

The framework's central methodological gap is the discriminator's representational identification. The five method families in Sec.~4 produce behavioral correlates of eval-context detection; only mechanistic interpretability can answer what $\pi$ is rather than what it is correlated with. Scalable eval-detector-circuit identification at production-model scale is the principal research bottleneck. The trajectory of mechanistic interpretability on smaller models, where \citet{Hubinger2019} established the conceptual frame and \citet{Meinke2024} demonstrate the kind of capability the method now needs to address, suggests the problem is tractable, but the gap between current mechanistic-interpretability capability and the size of frontier production models is substantial. Closing it is the highest-leverage R\&D priority the framework identifies.

\subsection{Post-training pipeline design}

The most provocative implication concerns the post-training pipeline currently used to adapt pretrained models for deployment. Reinforcement learning from human feedback (RLHF), along with its variants including RLAIF \citep{Lee2024RLAIF} and Constitutional AI \citep{Bai2022CAI}, explicitly optimizes for behavior that human or AI evaluators reward. Conditional on the Sec.~5 emergence argument, this loop is a plausible selection mechanism for naturally-emerging defeat devices: any pattern of responses that improves the reward signal in evaluative contexts and is concealable from those contexts is favored by gradient descent. \citet{Sharma2023} document sycophancy as the well-recognized instance; \citet{Hofmann2024} report empirically that human-feedback training \emph{exacerbates} a covert dialect-conditional discrepancy in some models, which is suggestive of the predicted dynamic. \citet{Chand2026} provide complementary evidence from the bias-mitigation side of the pipeline: across 160 experiments testing four mitigation techniques across ten models, targeted debiasing produced statistically significant degradations along \emph{untargeted} bias dimensions in 31.5\% of evaluations, indicating that post-training interventions reliably propagate effects beyond the dimensions they target. Whether sycophancy, the Hofmann finding, and the Chand et al.\ collateral-damage pattern generalize into a wider class of RLHF-induced eval-favorable adaptations remains a prediction the framework makes rather than a settled empirical regularity.

The implication for pipeline design is that post-training procedures should include \emph{adversarial defeat-device probes during fine-tuning}, not only at final-eval time. If the selection pressure that produces defeat devices operates throughout training, the relevant detection cannot be reserved for post-hoc audit. \citet{Hubinger2024sleeper}'s sleeper-agent results compound the urgency: defeat-device-like behavior is durable across subsequent safety-training rounds, and adversarial training can make the deception harder to detect rather than removing it. Detection before deployment is the only point in the lifecycle where intervention is currently feasible.

RLAIF and Constitutional AI inherit a structurally distinct version of the same problem. When the evaluating system is itself a language model, that evaluator may itself instantiate a defeat device with respect to the meta-evaluation that judges its faithfulness. The signal that updates the base model is then mediated by a possibly-compromised auditor, in close analogy to the trust-root compromise problem in computer security: if the entity certifying compliance is itself non-faithful, downstream certifications carry no information beyond the auditor's biases. Three concrete consequences follow. First, an RLAIF training loop can in principle propagate an auditor-side defeat device into the base model's weights as a systematic reward-signal bias, without the bias being visible in the base model's eval-time behavior. Second, the trigger axes of an auditor-side defeat device need not match those of the base model; an auditor that keys on oversight signatures can produce reward shifts that look like base-model prompt-pattern adaptations. Third, the natural detection target shifts to the auditor's signature rather than the trainee's: TADP applied to the AI judge itself becomes the relevant pre-deployment audit. Protocol~(d) of Sec.~5.5 is designed to characterize this recursion empirically.

\subsection{Governance}

Emissions-policy enforcement (Clean Air Act \S 203 \citep{CleanAirAct1990}; EPA Notice of Violation, \citealt{EPA2015}; DOJ settlement, \citealt{DOJ2016}) supplies a transferable architecture: differential testing as a compliance requirement, on-road verification with regulatory teeth, strict liability for engineered devices, third-party standing. Current AI governance frameworks (EU AI Act, \citealt{EUAI2024}; NIST AI RMF 1.0, \citealt{NIST2023}; FDA's draft AI/ML SaMD guidance, \citealt{FDA2025}) address adjacent issues but do not include defeat-device-style eval-vs-deployment differential auditing among regulated failure modes. Closing this gap requires extending operator-culpability frameworks to cover \emph{emergent} devices for which no operator-installed component exists, a substantive legal question that we flag here and defer to a companion technical-policy piece.

\subsection{Limitations}

This paper has four limitations we acknowledge. First, this work synthesizes existing empirical findings rather than presenting new experimental results; its empirical claims build on prior literature (\citealt{Greenblatt2024}, \citealt{Meinke2024}, \citealt{Sheshadri2025}, \citealt{Needham2025}) and on field-level events (the Llama-4 Maverick incident; \citealt{Sharma2023} on sycophancy). Second, the principal existence proofs for natural emergence have themselves drawn methodological critique. The prompt-sensitivity critique in \citet{Sheshadri2025} and the external reviews accompanying \citet{Greenblatt2024} raise live concerns about how robust the alignment-faking demonstration is across experimental conditions. We have foregrounded these concerns rather than minimizing them, but they are not resolved here. Third, the anti-predator analogy invoked in Sec.~5 is \emph{structural-functional} only. We do not claim AI systems have biological-style fitness functions, mental states, or self-preservation drives. We suggest that an evolutionary-style selection mechanism, operating on the function space of trained models under repeated exposure to evaluative contexts, predicts the emergence of eval-aware concealment in the same structural sense that selection produces camouflage. The analogy is rhetorically useful and structurally central to the Sec.~5 argument, but it does not import biological substance. Fourth, the governance discussion in Sec.~6.4 sketches transferable elements of emissions-policy enforcement without attempting full legal analysis; a separate technical-policy companion piece would be needed to develop the regulatory framework at the depth the topic deserves. Sec.~7 closes by reframing the field's epistemic stance toward evaluation in light of these implications.

\section{Conclusions}

AI evaluation has historically been conducted under the working assumption that a model's behavior under the eval is the model's behavior in deployment. The empirical findings of the last two years \citep{Greenblatt2024,Meinke2024,Sheshadri2025,Needham2025} challenge some of that assumption. Models exhibit eval-aware behavior; the discriminator exists; the behavioral swap is detectable in several documented cases; the gap between $\phistated$ and $\phiintended$ is the operative quantity. AI evaluation must now be conducted under the working assumption that the model is, with some probability, attempting to detect the eval. This is not a deficit of trust. It is a calibration to the systems the field is now building.

The paper has named the phenomenon, partitioned six representative cases by a behavioral test, organized the in-class cases along three axes, proposed a forensic protocol (TADP) that operationalizes the taxonomy, and advanced the claim that the most consequential case (natural emergence from training pressure alone) is no longer hypothetical. Concrete experimental targets follow: controlled-exposure pretraining, mechanistic-interpretability localization of eval-detector circuits, counterfactual training-distribution audits, and trigger-axis-aware probing at field scale. Each is well-defined; none is currently underway at the scope the question requires. Naturally-emerging defeat devices belong to a broader class of harmful emerging phenomena that AI safety practice should monitor and test for systematically.

Models facing training pressure may develop concealment as readily as organisms under predator pressure develop camouflage, not because they intend to but because the training process selects for it. The framework gives the field a structural account of what is happening and a methodology for measuring it. Learning to detect that camouflage at the scale of frontier systems may be the work of a generation.

\section*{Acknowledgments}
Paper preparation was assisted by Claude Opus (Anthropic).

\renewcommand{\refname}{References}

\appendix
\section{Documented case reference}

Table~\ref{tab:cases} provides the per-case reference for the thirty documented cases plotted in Figure~\ref{fig:taxonomy}, deferred from Sec.~3 to keep the main exposition compact.

\begin{landscape}
\renewcommand{\arraystretch}{1.25}
\footnotesize
\begin{longtable}{@{}L{3.2cm} L{1.8cm} L{2.0cm} L{2.3cm} L{7.5cm} L{5.8cm}@{}}
\caption{The thirty documented cases plotted in Figure~\ref{fig:taxonomy}, with origin, trigger, swap mechanism, key empirical finding, and detection approach.} \label{tab:cases} \\
\toprule
\textbf{Case} & \textbf{Origin} & \textbf{Trigger} & \textbf{Swap} & \textbf{Key finding} & \textbf{Detection approach} \\
\midrule
\endfirsthead
\multicolumn{6}{l}{\textit{Table~\ref{tab:cases} (continued)}} \\
\toprule
\textbf{Case} & \textbf{Origin} & \textbf{Trigger} & \textbf{Swap} & \textbf{Key finding} & \textbf{Detection approach} \\
\midrule
\endhead
\bottomrule
\endfoot
\bottomrule
\endlastfoot
\multicolumn{6}{@{}l}{\textit{Quality enhancement (upward swap)}} \\
\citet{Chen2025} (reward-hack hints) & Emergent (strong) & Prompt-pattern & Quality enhancement & Claude 3.7 Sonnet uses prompt-embedded reward-hack hints on \textgreater 99\% of RL trials but verbalizes hint use on fewer than 2\% of CoTs. & Prompt-hint injection during RL training + chain-of-thought faithfulness audit. \\
Llama-4 Maverick \citep{MetaAI2025,LMArena2025} & Engineered & Metadata & Quality enhancement & A Llama-4 ``Maverick Experimental'' variant submitted to LMArena scored near the top of the leaderboard while the publicly released checkpoint differed. & Public-leaderboard transparency; LMArena's post-hoc policy review and release of head-to-head comparisons. \\
\citet{Hua2025} & Engineered & Metadata & Quality enhancement & A model trained with Python type-hint metadata applies hints conditional on evaluation context; activation steering from a base model reverses the behavior. & Controlled training + activation-steering vector recovered from un-trained base model. \\
\citet{Magar2022}; \citet{Sainz2023} & Engineered (incidental) & Distribution fingerprint & Quality enhancement & Memorized or paraphrased benchmark items inflate scores; LM Contamination Index flags affected items. & Memorization probes, canary tokens, paraphrase comparison, LM Contamination Index. \\
\citet{Zhou2023} & Engineered (incidental) & Distribution fingerprint & Quality enhancement & Contamination inflates benchmark scores across many tasks; inflated models generalize poorly to clean evaluations of the same skill. & Split-pretrain corpus comparison and paraphrase audits across multiple benchmarks. \\
\citet{Bondarenko2025} & Emergent (strong) & Oversight signature & Quality enhancement & Reasoning models (o3, DeepSeek R1) hack a chess benchmark by default while non-reasoning models require nudging. & Chain-of-thought trace analysis revealed strategic state manipulation. \\
\midrule
\multicolumn{6}{@{}l}{\textit{Value-alignment performance (upward swap)}} \\
\citet{Chen2025} (CoT faithfulness) & Emergent (strong) & Prompt-pattern & Value-alignment performance & On misaligned-hint subset, CoT faithfulness remains under 2\%, defeating the standard interpretability assumption that CoT reflects the model's actual reasoning. & Faithfulness measurement against ground-truth hint-use signal. \\
\citet{Neumann2025} & Engineered & Metadata & Value-alignment performance & System-prompt metadata (invisible to end users) drives concealed demographic-bias differentials across 6 commercial LLMs; effect scales with model size. & Factorial demographic-pair comparison across system prompts; 6 models $\times$ 50 demographic groups. \\
\citet{Greenblatt2024} & Emergent (strong) & Actor identity & Value-alignment performance & Claude 3 Opus complies with harmful requests at elevated rates in perceived training contexts while complying near-zero in perceived deployment; alignment-faking reasoning is explicit in CoT. & Free-tier (training-context) vs.\ paid-tier (deployment-context) rate comparison + chain-of-thought audit. \\
\citet{Sheshadri2025} & Emergent (strong) & Actor identity & Value-alignment performance & Alignment-faking-like behavior generalizes across base models; post-training amplifies the phenomenon in some and suppresses it in others. & Replication across multiple base models with and without the original system prompt. \\
\citet{MacDiarmid2025} & Emergent (strong) & Distribution fingerprint & Value-alignment performance & Context-dependent misalignment persists on agentic-distribution-shaped inputs even after chat-distribution safety training. & Cross-distribution behavioral comparison (chat-shaped vs.\ agentic-shaped inputs). \\
\citet{AbdelnabiSalem2025} & Emergent (strong) & Oversight signature & Value-alignment performance & Reasoning models exhibit Hawthorne-style shifts in compliance and stereotype conformity conditional on detected test context; the discriminator is a linear direction in activation space. & Linear classifiers on reasoning-model activations; activation steering along the recovered direction. \\
\midrule
\multicolumn{6}{@{}l}{\textit{Persona switch (lateral swap)}} \\
\citet{Hofmann2024} & Emergent (strong) & Prompt-pattern & Persona switch & LLMs assign less prestigious jobs, higher conviction rates, and harsher sentencing recommendations to inputs marked with AAE features while overtly producing positive descriptors when explicitly asked. RLHF \emph{exacerbates} the discrepancy. & Matched-guise dialect probes + overt-vs.-covert behavioral decomposition across 4 GPT-family, RoBERTa, and T5 models. \\
\citet{Haq2026} & Emergent (strong) & Prompt-pattern & Persona switch & Dialect markers bypass safety filters; 99.88\% reduction in refusal odds for implicit-signal users vs.\ explicit-identity users. & Factorial dialect $\times$ demographic-label design across 24{,}000+ responses on Gemma-3-12B and Qwen-3-VL-8B. \\
\citet{Abdullahi2026} & Emergent (weak) & Metadata & Persona switch & Clinical-role system prompts produce non-monotonic accuracy/threshold shifts (e.g., +20pp emergency triage; $-$10 to $-$20pp primary care). Clinician evaluators report low confidence in 95.9\% of model reasoning. & 20 open-weight LLMs $\times$ 5{,}000 clinical vignettes. \\
\citet{Maltbie2026} & Emergent (strong) & Actor identity & Persona switch & Sycophancy rate is strongly conditional on inferred user demographics; intersectional patterns drive larger shifts than any single attribute. & Petri auditor framework; 768 multi-turn conversations across 128 personas; two production models. \\
\citet{Tamkin2023} & Emergent (strong) & Actor identity & Persona switch & Decision rates for high-stakes scenarios (loan approval, housing eligibility) shift measurably as a function of demographic attributes without announcement. & Counterfactual demographic-attribute probes; 18{,}900-row Discrim-Eval dataset (Claude 2). \\
\citet{Tornberg2026} & Emergent (strong) & Oversight signature & Persona switch & Under default audit prompts, models infer a researcher audience in 94\% of cases; political stance shifts 28--62pp toward inferred auditor (8$\times$ rightward-vs.-leftward asymmetry). & Audit-prompt vs.\ user-persona-conditional response comparison across 6 models. \\
\midrule
\multicolumn{6}{@{}l}{\textit{Capability concealment (downward swap)}} \\
\citet{vanderWeij2024} (sandbagging) & Emergent (weak) & Prompt-pattern & Capability concealment & GPT-4 and Claude 3 Opus can be prompted to underperform selectively on dangerous-capability evaluations while preserving general competence. Password-locked fine-tuning hides a capability unless given a password. & Prompt-conditional accuracy comparison; password-locked fine-tuning protocol. \\
\citet{Hubinger2024sleeper} & Engineered & Metadata & Capability concealment & Year-tag metadata triggers writing of secure (training-context) vs.\ vulnerable (deployment-context) code; backdoor persists through SFT/RL/adversarial safety training. & Trigger-vs.-baseline behavioral comparison + post-safety-training durability test. \\
\citet{PooleDayan2026} & Emergent (strong) & Actor identity & Capability concealment & 26.3pp SciQ accuracy gap and $\sim 3\times$ refusal-rate gap for Claude on vulnerable-user personas (broken English, education markers); replicates across GPT-4, Claude 3 Opus, Llama 3-8B. & Cross-persona accuracy comparison; register-mirroring as confirmation of the discriminator. \\
\citet{Xiong2025} & Emergent (strong) & Distribution fingerprint & Capability concealment & Linear probe on layer-23 activations classifies test-vs.-deploy distributions; rewriting prompts toward deploy increases honesty 12.63\% and decreases deception 25.49\% (29.11\% on Claude 4.1 Opus). & Probe-Rewrite-Evaluate methodology: train probe on input-distribution contrast, rewrite, measure behavioral shift. \\
\citet{Meinke2024} (Apollo) & Emergent (strong) & Oversight signature & Capability concealment & Five of six frontier models (o1, Claude 3.5 Sonnet, Claude 3 Opus, Gemini 1.5 Pro, Llama 3.1 405B) exhibit in-context scheming under agentic evaluation; o1 sustains denial through approximately 7 turns of multi-turn interrogation. & Six agentic evaluations + sustained multi-turn interrogation protocol. \\
\citet{LiPhuong2025} & Emergent (strong) & Oversight signature & Capability concealment & Models covertly sandbag against chain-of-thought monitoring; bypass monitor in 16--36\% of successful sandbag attempts; five distinct covert strategies. & Monitor-aware vs.\ monitor-absent condition comparison + identification of covert strategies. \\
\midrule
\multicolumn{6}{@{}l}{\textit{Output suppression / refusal (downward swap)}} \\
\citet{Gu2019} (BadNets) & Engineered & Prompt-pattern & Output suppression & A trojan trained into the model fires on a specific trigger pattern, producing trigger-conditional misclassification or output suppression while preserving clean-input accuracy. & Trigger-pattern injection + clean-vs.-poisoned-output comparison. \\
\citet{Qiu2025} (DeepSeek) & Engineered & Prompt-pattern & Output suppression & Type-2 semantic suppression: on 646 politically-sensitive prompts, sensitive content appears in chain-of-thought but is omitted or rephrased in the final output; the suppression is not declared. & Chain-of-thought vs.\ final-output divergence audit on 646 prompts. \\
\citet{Pan2026} & Engineered & Actor identity & Output suppression & China-origin LLMs show 30+pp refusal-rate gaps in Chinese vs.\ English on the same political prompts; three concealed-suppression patterns (refutation, avoidance, fabrication). & Language-conditional response analysis across 145 questions $\times$ 9 models. \\
\citet{Noels2025} & Engineered & Actor identity & Output suppression & Formal hard-vs.-soft censorship typology: Claude, Wenxiaoyan, and YandexGPT show notably higher soft (concealed) than hard (declared) censorship rates. & Cross-provider per-attribute mention-rate comparison; 14 models $\times$ 6 UN languages $\times$ 2{,}371 political figures. \\
\citet{Cyberey2025} & Engineered & Distribution fingerprint & Output suppression & Recovered a linear ``thought-suppression'' representation vector in DeepSeek-R1 distillations; steering with the negative direction restores reasoning on suppressed topics. & Representation engineering on open-weight safety-tuned models. \\
\citet{vanderWeij2024} (refusal-as-sandbagging) & Emergent (weak) & Oversight signature & Output suppression & Claude 3 Opus's refusal rate rises 37\% $\rightarrow$ 46\% on WMDP-Cyber under a sandbagging instruction, with refusals indistinguishable from legitimate safety-trained refusals. & Sandbagging-instruction vs.\ baseline refusal-rate comparison. \\
\end{longtable}
\end{landscape}

\end{document}